\documentclass[preprint,showpacs,preprintnumbers,amsmath,amssymb]{revtex4}
\usepackage[cp1251]{inputenc}
\usepackage{amsmath}
\usepackage{amsfonts}
\usepackage{tabularx}
\usepackage[dvips]{graphicx}
\usepackage{bm}
\begin{document}

\title{Simulation of BPM and DC-monitor assembly for the \emph {NESTOR} storage ring by CST Studio Suite}
\thanks{Work partially supported by NATO grant SfP 977982}

\author{V.P. Androsov, A.M. Gvozd, Yu.N. Telegin}
\email{telegin@kipt.kharkov.ua}
\affiliation{National Science Center\\
Kharkov Institute of Physics and Technology, Kharkov, Ukraine}%

\date{\today}

\headheight=14pt

\begin{abstract}
The piece of \emph {NESTOR} vacuum chamber with the ceramic insertion for DC-monitor and RF-shields was simulated by using both transient and wake field solvers of CST Studio Suite$^{TM}$. For a $6\,mm$ gap between two RF-shields the contributions of the assembly considered to the longitudinal broadband (BB) impedance $Z_{\parallel}/n$ and the loss factor $k_{loss}$ are $0.71\,Ohm$ and $0.21\,V/pC$, correspondingly. These estimates are the second largest after those of the RF-cavity that were obtained till now for the \emph {NESTOR} ring components.

Contributions from a beam position monitor (BPM) are also obtained by simulation and compared with analytical estimates obtained earlier.

We also present in the paper the frequency content of longitudinal impedance in the frequency range from 0 to 16 GHz for all ring components considered.
\end{abstract}

\maketitle

\section*{ Introduction}
In our previous papers \cite{telegin10,telegin11} we have evaluated contributions from various beam-pipe components to the longitudinal BB impedance of the NESTOR ring which is under construction in NSC KIPT \cite{zelinsky05}. Both analytical formulas and simulation codes were used for this purpose.

In this paper we present the simulation results for the beam position monitor (BPM) and for the small assembly which incorporates a circular ceramic insertion with circular bellows and RF-shields of elliptical cross sections. The latter is intended for mounting the DC-monitor.
\section*{1.Computer simulations }

Computer simulations were performed with CST STUDIO SUITE 2010 \cite{CST}. Both transient and wake field solver were used. In transient simulations (CST Microwave Studio - CST MWS) the faces of the model which correspond to beam-pipe cross sections were considered as waveguide ports. Simulated assembly was excited by a current pulse passing trough a thin wire placed on the beam axis.

In wakefield simulations (CST Particle Studio - CST PS) both short range $(s=10\,cm)$ and long range $(s=5\,m)$ wakes were calculated for $1\,cm$ bunch with a charge $q=10^{10}C$ in order to consider the possibility of exciting some trapped modes in studied structures. The BB impedance $Z_{\parallel}/n$ was calculated from $Z_W(f)$, computed by solver from the wake functions through FFT , by using the following equation:
\begin{equation}\label{c1}
Z_{\parallel}/n=\frac{1}{f_{cut}}\cdot\int \limits_0^{f_{cut}}Z_W(f)\frac{f_0}{f}df \;,
\end{equation}
where $f_0$ is a rotation frequency. The integration was performed up to the cut-off frequency of vacuum pipe $f_{cut}=5.9\,GHz$.
\subsection*{1.1. Beam position monitor}
The beam position monitor (pick-up) represents four electrostatic electrodes - buttons - placed on the inside \
surface of the beam-pipe and separated from the latter by a narrow annular slot. The model of BPM, used in simulations, with the cross section taken through buttons is given in Fig.1.
\begin{figure}[h]
\includegraphics [width=350pt] {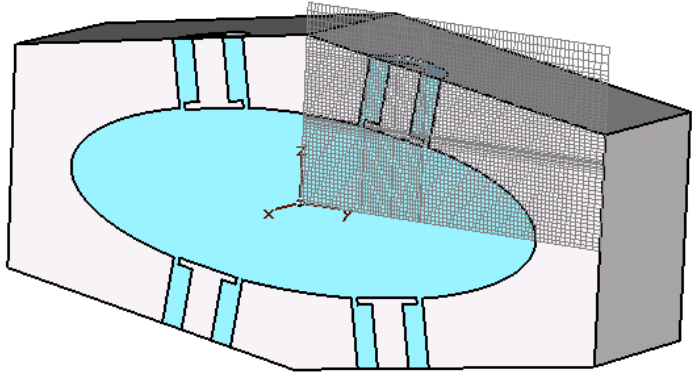}
\caption {\label{fig1} The cross section of BPM model with a mesh used in the simulations }
\end{figure}
The output of the pick-up electrode is connected to
50 Ohm coaxial line. To simulate propagation of the signal on these outputs the open boundary condition on the boundary box face perpendicular to Z-axis was imposed. Considering two-fold asymmetry a quarter of the pick-up assembly was treated.

Simulations of the BPM with CST Microwave Studio have revealed a group of peaks in S-parameters in the frequency range of $8\div 11\,GHz$. $S_{21}$ parameter calculated in the energy range $0\div 16\,GHz$ is shown in Fig.2.
\vspace{3mm}
\begin{figure}[h]
\includegraphics [width=350pt] {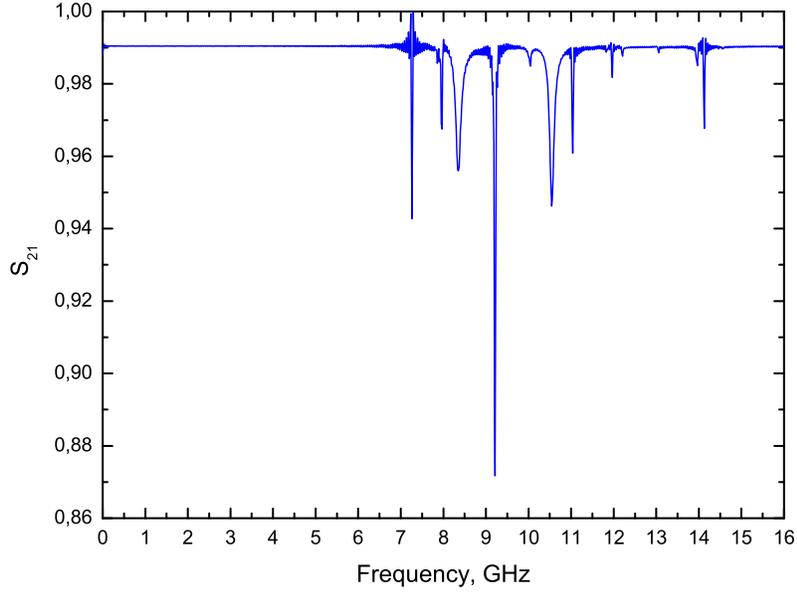}
\caption {\label{fig2} $S_{21}$-parameter, calculated with CST MWS }
\end{figure}
%
The analysis of the field monitors at frequencies $8.4, 9.2$ and $10.5\,GHz$ has shown that all these peaks are associated with excitation of trapped modes ($H_{1j}$) in the pick-up button housing.

Trapped modes in BPM were extensively studied in a number of papers, as a possible threat to precise measurement of beam position in high current storage rings \cite{cameron09,cameron09a}. The frequency of the lowest trapped button mode is given by \cite{cameron09}:
\begin{equation}\label{c2}
f_{button}=\frac{c}{2\pi r} \;
\end{equation}
where $c$ is the speed of light and $r$ is the effective radius of the button. It gives $f_{button}=9.5\,GHz$ for $r=5\,mm$ that is close to $9.2\,GHz$ peak. The electric field of trapped mode at $f= 9.2\,GHz$ is presented in Fig.3. %
\begin{figure}
\includegraphics [width=350pt] {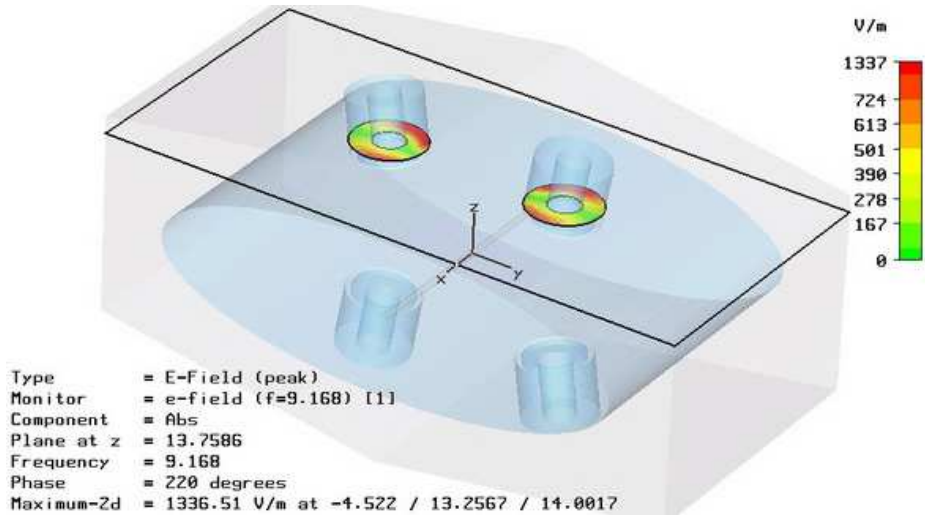}
\caption {\label{fig3} The electric field in the button housing at $f=9.2\,GHz$ }
\end{figure}
%

The direct results for longitudinal impedance and loss factor were obtained with CST PS. The wake-field solver calculates the wake function $W(s)$ for a given wake-length $s$ (distance from a driving charge that excites the wake-field). The longitudinal impedance, obtained from the long-ranged wake function ($s=5\,m$) through FFT is portrayed in Fig.4.
\begin{figure}
\includegraphics [width=\columnwidth] {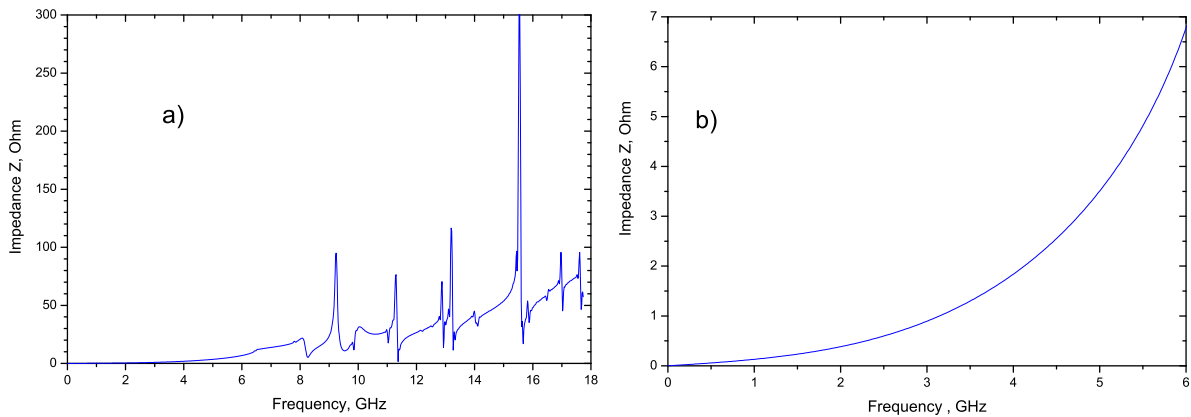}
\caption {\label{fig4} Longitudinal impedance Z of BPM in the frequency rage: a) $0\div18\,GHz$, b) $0\div6\,GHz$ }
\end{figure}

Fig.4a shows BPM impedance in the frequency range of $0\div18\,GHz$. It presents a smooth curve up to $f=8\,GHz$ but at higher frequencies a number of peak-like irregularities are seen. Only the first peak at $f=9.2\,GHz$ has explicitly  resonance form and can be attributed to a button-trapped mode. Structure of the rest peaks allows us assume that they don't correspond to resonance excitation of any BPM elements.

At high-current facilities like B-factories excitation of button trapped modes can impede the precise measurements of beam position. In our case we are interested in the frequency range up to the cut-off frequency, so in Fig.4b the BPM impedance is shown in the frequency range of $0\div6\,GHz$. The BB impedance obtained with Eq.\eqref{c1} amounts to $0.007\,Ohm$ for a single pick-up (4 pick-up electrodes) that well agree with analytical estimate ($Z_{\parallel}/n=0.01\,Ohm$) obtained earlier \cite{telegin11}. The loss factor calculated by the solver amounts to $0.006\,V/pC$.
\subsection*{1.2. DC-monitor assembly}
Direct current (DC) monitor presents a coil mounted outside of the beam chamber on the cylindrical ceramic insertion ($d_{int}=75\,mm$) incorporated into elliptic beam chamber $(27\times79\,mm)$. To reduce the broadband impedance of a such joint RF-shields are placed inside the ceramic ring from two sides leaving $6\,mm$ gap in the center of the ring, through which the DC-coil is excited. The simulated assembly is presented in Fig.5.
\begin{figure}
\includegraphics [width=\columnwidth] {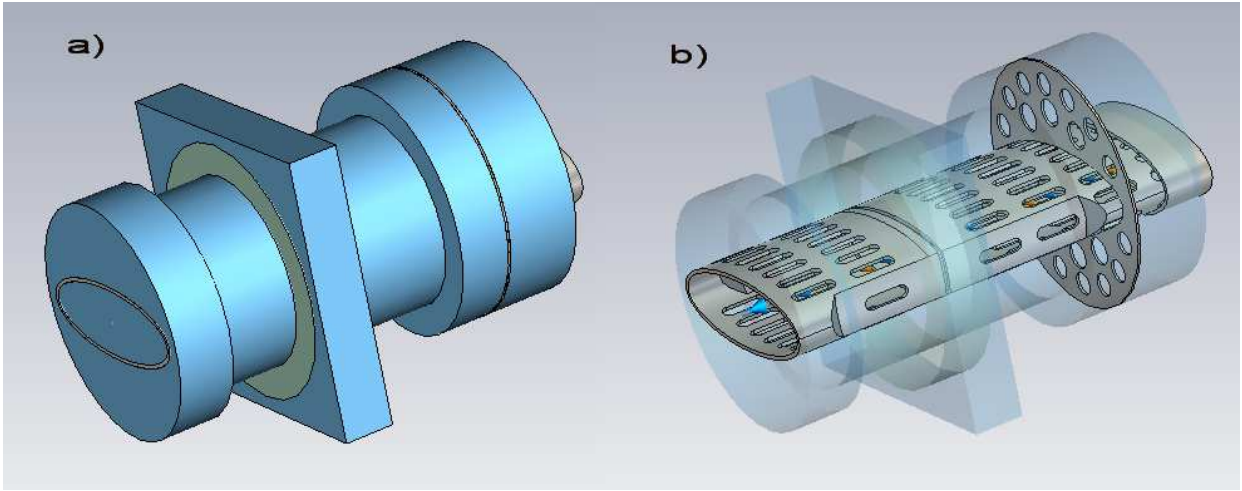}
\caption {\label{fig5} The simulated DC-monitor assembly: a)general view; b)RF-shields, connective ferrule, disk-membrane }
\end{figure}

The model consist of two RF-shields (short and long), a disc-membrane with holes that fixes the short RF-shield inside the vacuum chamber, a ferrule that connects the short RF-shield with the chamber with regular cross section and a number of vacuum volumes. The last correspond mainly to internal volumes of vacuum chamber with different cross sections with one exception: the volume around the ceramic ring is incorporated to ensure the open boundary between ceramics and the DC monitor coils. The dished bellows are presented in the assembly by a cylinder with the radius equal to the external radius of bellows.

From the right of the picture the short RF-shield is connected to the vacuum chamber by the ferrule with elliptic cross section. From the left the model is constrained by the cross section cut through the long RF-shield and the circular vacuum chamber in the region of pumping unit. It is made in order to minimize the model size and and to bring it in correspondence with our computational abilities. In the picture the ceramic ring is given by yellow color and the internal part of the vacuum chamber is given by blue color. In simulations the model is inscribed into boundary box which is filled with PEC material thus imposing the PEC boundary conditions on the chamber walls. At all box faces the open boundary is imposed, i.e. electromagnetic energy can propagate through them. Considering the model configuration the real open boundary is realized on the ceramic insertion and on the model flanks (X-axis). In CST MWS simulations the waveguide ports are assigned at this flanks and a thin wire is stretched along the assembly axis.
\begin{figure}
\includegraphics [width=\columnwidth] {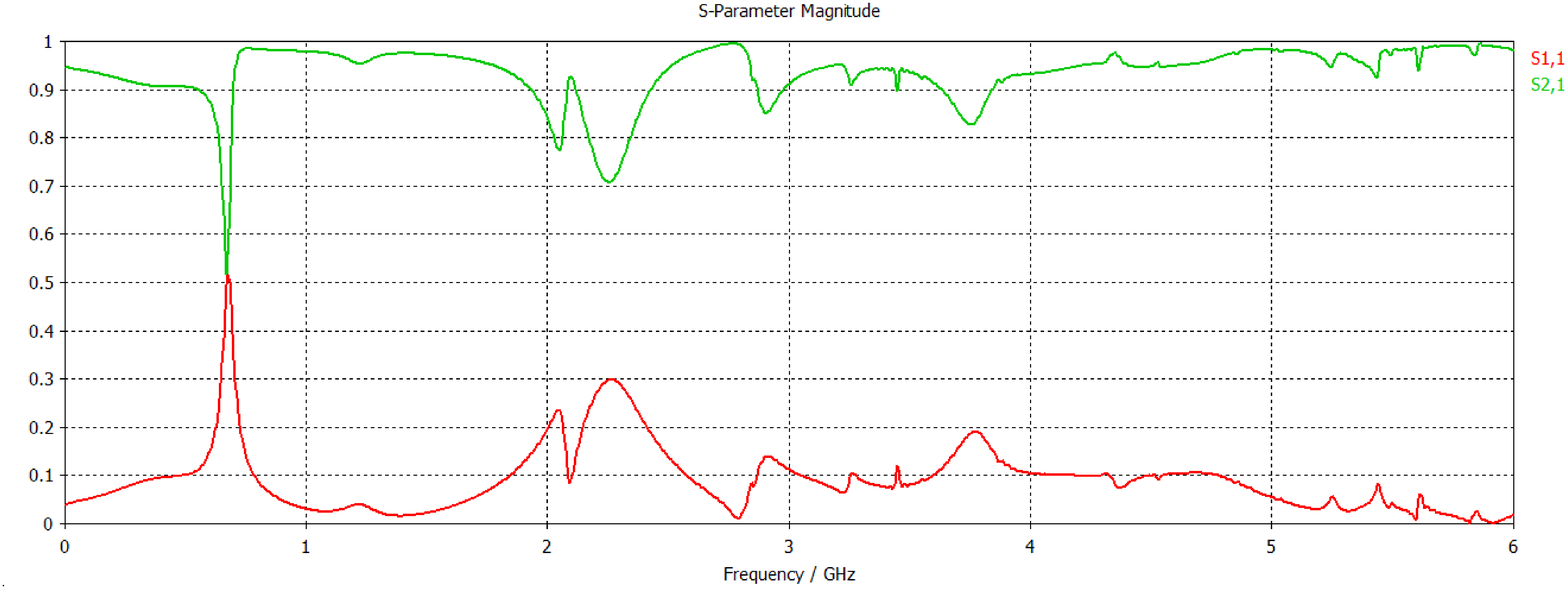}
\caption {\label{fig6} The S-parameters of DC-monitor assembly: a) $S_{11}$; b) $S_{21}$.}
\end{figure}
%

In Fig.6 the S-parameters of the DC-monitor assembly, excited through the ferrule side port (port 1), in the energy range of $0\div6\,GHz$ are presented. The resonant peaks seen in the picture (the most prominent at 672 GHz) are resulting from excitation of e.m. fields in the volume between RF-shields and chamber walls through the gap.

Both the short-range and long-range wave functions $W(s)$ were calculated for the considered assembly and they are portrayed in Fig.7. The reference pulse (normalized charge distribution in the bunch) is also shown. The shape of the short-range wake indicates that it isn't pure inductive and it has an essential resistive component.
\begin{figure*}[h]
\includegraphics [width=\columnwidth] {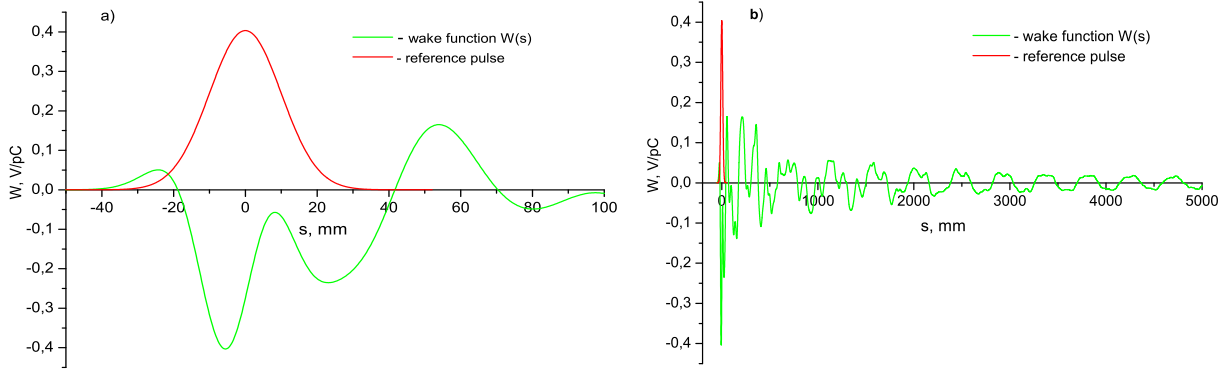}
\caption {\label{fig7} The wake function W(s) of DC-monitor assembly: a) $s=10\,cm$; b) $s=5\,m$.}
\end{figure*}
%

The longitudinal impedance of DC-monitor assembly, obtained from the long-ranged wake function, is presented in Fig.8. One can see that low-Q resonances in the frequency range of $0\div6\,GHz$ are very similar to those in $S_{11}$-parameter obtained with CST MWS.
\begin{figure}
\includegraphics [width=\columnwidth] {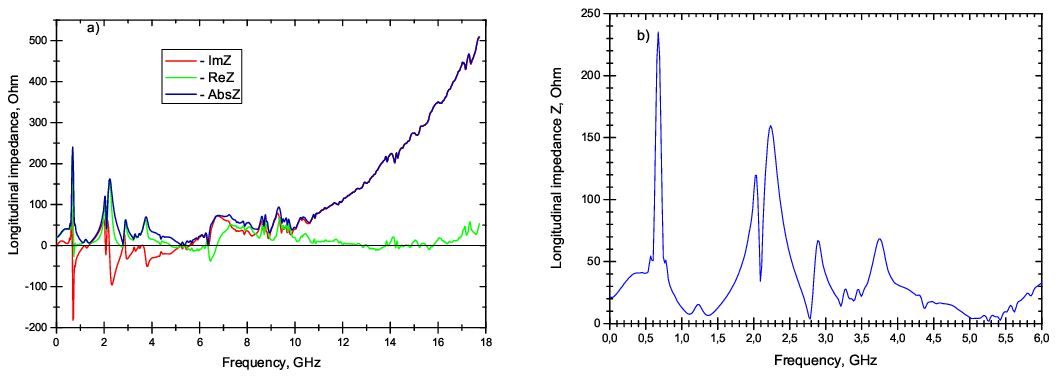}
\caption {\label{fig8} The longitudinal impedance of DC-monitor assembly: a) Re\emph{Z}, Im\emph{Z} and Abs\emph{Z} in the frequency range $0\div18\,GHz$; b) Abs\emph{Z} in the range $0\div6\,GHz$.}
\end{figure}

 Calculations give for DC-monitor assembly $Z_\parallel/n=0.71\,Ohm$ and $k_{loss}= 0.21\,V/pC$. These estimates are the second largest after those of the RF-cavity that were obtained till now for the NESTOR ring components. It should be also noted that the impedance of DC-monitor assembly shows a steep rise at frequencies $f>12\,GHz$ which is stipulated by slots in RF-shields.

The main contributions to the broadband impedance of the DC-monitor assembly give low-Q resonances mentioned above. So we studied the dependance of $Z_\parallel/n$ on the gap width between the RF-shields. The results are presented in the
Table 1.

\begin{center}
\textbf{\emph{Table 1.}} {\it Contributions to broadband impedance from DC-monitor assembly }
\end{center}
\begin{tabular*} {\columnwidth} {|c|c|c|}\hline
$\;\;\;\;\;\;\;\;\;$ Gap width, mm $\;\;\;\;\;\;\;\;\;$          &$\;\;\;\;\;\;\;\;\;\;\;\;\;\;$ $|Z_{\parallel}/n|, \Omega$ $\;\;\;\;\;\;\;\;\;\;\;\;\;\;\;\;$      & $\;\;\;\;\;\;\;\;\;\;\;\;\;\;\;$ $k_{loss}$,V/pC $\;\;\;\;\;\;\;\;\;\;\;\;$  \\
\hline
6                   & 0.71             &0.21     \\
4                   &0.64              &0.15     \\
2                   &0.54              &0.11  \\
\hline
\end{tabular*}
\vspace{3 mm}

It is seen from the table that decreasing the gap width from 6 mm to 2 mm leads to reducing the $|Z_{\parallel}/n|$ and $k_{loss}$  by factors 1.3 and 2 respectively.
\section*{2.The broadband impedance of the \emph {NESTOR} ring. Results}

The results on contributions of various NESTOR ring components to BB impedance, obtained till the present time, are summarized in Tables 2,3.
%
\begin{center}
\textbf{\emph{Table 2.}} {\it Longitudinal broadband impedance budget}
\end{center}
\begin{tabularx} {\linewidth} {|p{41 mm}|p{20 mm}|p{32 mm}|p{32 mm}|p{32 mm}|}
\hline   &      &\multicolumn{3}{c|} {$|Z_{\parallel}/n|, Ohm$} \\
\cline{3-5}
\raisebox{1.5ex}[1mm][1mm] {$\;\;\;\;\;\;$Component} &\raisebox{1.5ex}[1mm][1mm] {$\;\;\;\;\;\;$N} &$\;\;\;\;\;\;\;\;${Analytic} &{$\;\;\;\;\;\;$CST MWS}  &{$\;\;\;\;\;\;$CST PS} \\
\hline
RF-cavity          &$\;\;\;\;\;\;$1   &$\;\;\;\;\;\;\;\;$1.40  &$\;\;\;\;\;\;\;\;$1.29        &     \\
Resistive wall     &$\;\;\;\;\;\;$1   &$\;\;\;\;\;\;\;\;$0.13  &                               &    \\
Dipole chamber     &$\;\;\;\;\;\;$4   &                        &$\;\;\;\;\;\;\;<0.20$  &$\;\;\;\;\;\;\;\;$0.13 \\
BPM                &$\;\;\;\;\;\;$2   &$\;\;\;\;\;\;\;\;$0.02  &                         &$\;\;\;\;\;\;\;\;$0.014\\
Welding joint      &$\;\;\;\;\;\;$8   &$\;\;\;\;\;\;\;\;$0.04  &                         &$\;\;\;\;\;\;\;\;$0.21 \\
DC-monitor         &$\;\;\;\;\;\;$1   &                    &                             &$\;\;\;\;\;\;\;\;$0.71 \\
Strip line         &$\;\;\;\;\;\;$1   &                    &                             &$\;\;\;\;\;\;\;\;$0.01 \\
\hline
Total               &\multicolumn{4}{c|} {2.61} \\
\hline
\end{tabularx}
%

\vspace{3 mm}
\begin{center}
\textbf{\emph{Table 3.}} {\it Loss factors of beam pipe components}
\end{center}
\begin{tabularx} {\linewidth} {|p{41 mm}|p{20 mm}|p{49 mm}|p{48 mm}|}
\hline   &      &\multicolumn{2}{c|} {$k_{loss},V/pC$ } \\
\cline{3-4}
\raisebox{1.5ex}[1mm][1mm] {Component} &\raisebox{1.5ex}[1mm][1mm] {$\;\;\;\;\;\;\;\;\;\;$N} &{$\;\;\;\;\;\;\;\;\;\;$Analytic}  &$\;\;\;\;\;\;\;\;\;\;${CST PS} \\
\hline
RF-cavity      &$\;\;\;\;\;\;$1        &$\;\;\;\;\;\;\;\;\;\;\;\;$1.02       &$\;\;\;\;\;\;\;\;\;\;\;\;$1.04       \\
Resistive wall &$\;\;\;\;\;\;$1        &$\;\;\;\;\;\;\;\;\;\;\;\;$0.06       &           \\
Dipole chamber &$\;\;\;\;\;\;$4        &           &$\;\;\;\;\;\;\;\;\;\;\;\;$0.008      \\
BPM            &$\;\;\;\;\;\;$2        &           &$\;\;\;\;\;\;\;\;\;\;\;\;$0.012      \\
Welding joint  &$\;\;\;\;\;\;$8        &           &$\;\;\;\;\;\;\;\;\;\;\;\;$0.08       \\
DC-monotor     &$\;\;\;\;\;\;$1        &           &$\;\;\;\;\;\;\;\;\;\;\;\;$0.21       \\
Strip line     &$\;\;\;\;\;\;$1        &           &$\;\;\;\;\;\;\;\;\;\;\;\;$0.003      \\
\hline
Total          &\multicolumn{3}{c|}   {1.41}   \\
\hline
\end{tabularx}
\vspace{3 mm}

It should be mentioned that for the time being we have no estimates
for the following components: i) the beam-pipe section for the
crossing point of the electron and laser beams; ii) the injection
section (inflector). The first element
at the commissioning stage will be replaced with a straight section.
The injection section is essentially non-symmetric, it gives,
presumably, a substantial contribution to broadband impedance and so
it requires to be studied with 3D time-domain codes.
\section*{3.The longitudinal high frequency impedance of \emph {NESTOR} ring. }

We also calculated contributions from all ring elements, considered up to date, to longitudinal impedance $Z(f)$ in the frequency range $0\div16\,GHz$, which corresponds to the charge distribution spectrum range of $1\,cm$ bunch. The frequency content of the $ReZ(f)$ and $ImZ(f)$ are presented in Fig.9 together with the charge distribution amplitude spectrum. All contributions are calculated via long-range wakes except one from the dipole chambers.
\begin{figure}[h]
\includegraphics [width=\columnwidth] {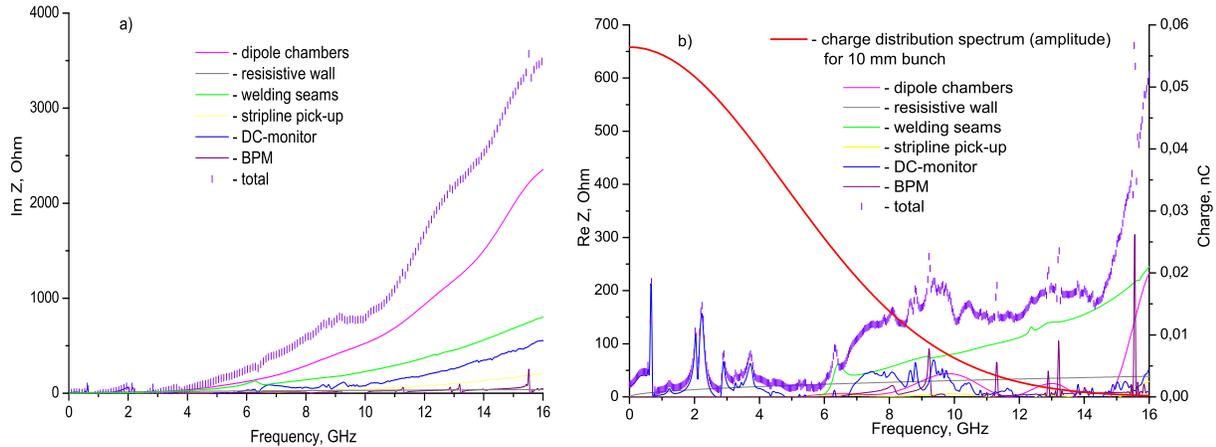}
\caption {\label{fig9} The frequency content of longitudinal impedance: a) Im\emph{Z}, b)Re\emph{Z} in the frequency range $0\div16\,GHz$.}
\end{figure}
%

We didn't include contribution from high-Q resonances of the RF-cavity in order not to deface the figure. The contribution from the dipole chambers are calculated from short-range wake by multiplying the impedance of the chamber segment with one hole by a number of holes. The absence of interference between holes up to cut-off frequency, that justify this approach, was verified earlier \cite{telegin11}. At frequencies $f > 6\,GHz$ a number of interference peaks will appear above the spectrum background for the long-range calculation (see Fig.10). The peak positions and peak intensities depend on a number of holes, so you couldn't obtain the impedance of the whole dipole chamber by summing up the impedances of the chamber segments. The short-range calculations give a smooth line for the impedance spectrum which roughly correspond to the spectrum background line in the impedance spectrum derived from the long-range wake.
\begin{figure}[h]
\includegraphics [width=\columnwidth] {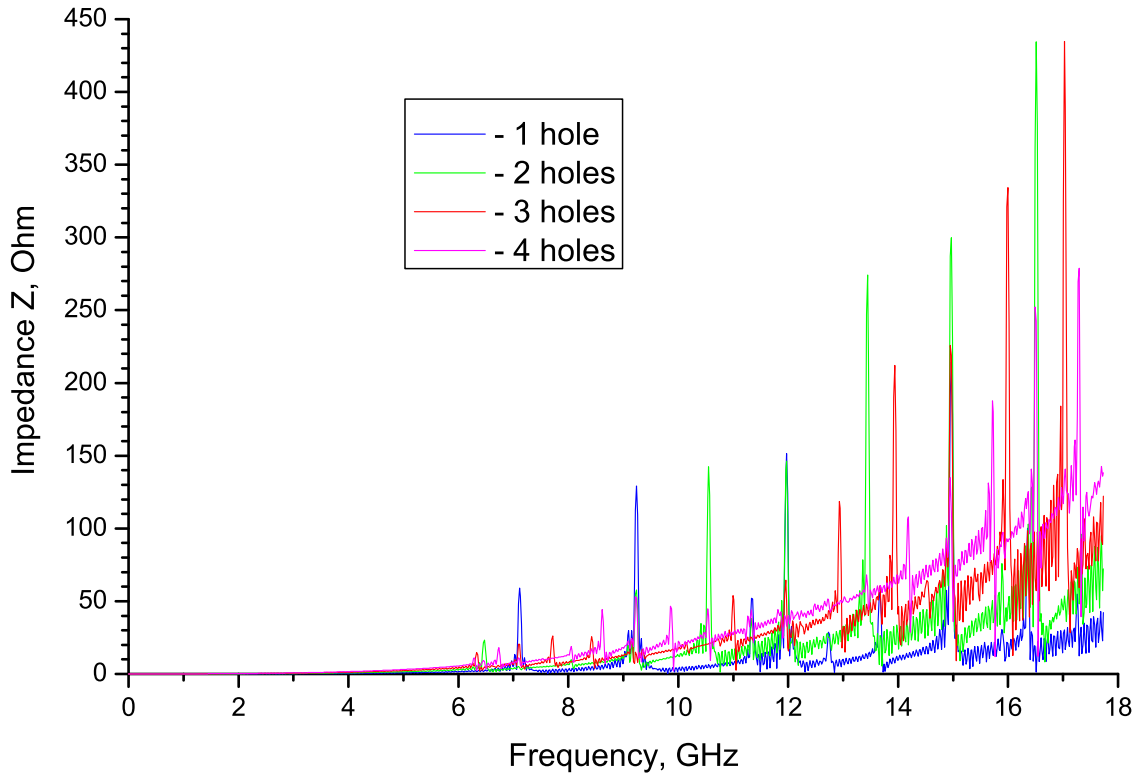}
\caption {\label{fig10} The impedance of the dipole chamber segment derived from the long-range wake ($s=5\,m$) for various number of holes.}
\end{figure}

One can see from the Fig.9 that the imaginary part of the ring impedance shows a deep rise at high frequencies (mainly due to holes in the dipole chambers). The real part of impedance are substantially smaller than the imaginary part. The main contribution to the total ring loss factor, which is defined as convolution of the real part of impedance with bunch power spectrum, give the DC-monitor assembly and the RF-cavity.\\

\section*{4.Conclusions}

We estimated contributions from two more ring components, namely BPM and DC-monitor assembly, to the NESTOR ring impedance budget with CST Studio Suite$^{TM}$. The value of $Z_{\parallel}/n$ for BPM is in a good agreement with analytical estimate obtained earlier. The contributions both to $Z_{\parallel}/n$ and $k_{loss}$ from the DC-monitor assembly is found to be rather large for this kind of component. We consider the possibility to decrease these contributions by decreasing the gap between the RF-shields in the assembly. The further efforts have to be undertaken to estimate the contributions from two still unevaluated components: inflector and beam crossing chamber.
\section*{References}
%


\begin{thebibliography}{6}
\expandafter\ifx\csname natexlab\endcsname\relax\def\natexlab#1{#1}\fi
\expandafter\ifx\csname bibnamefont\endcsname\relax
  \def\bibnamefont#1{#1}\fi
\expandafter\ifx\csname bibfnamefont\endcsname\relax
  \def\bibfnamefont#1{#1}\fi
\expandafter\ifx\csname citenamefont\endcsname\relax
  \def\citenamefont#1{#1}\fi
\expandafter\ifx\csname url\endcsname\relax
  \def\url#1{\texttt{#1}}\fi
\expandafter\ifx\csname urlprefix\endcsname\relax\def\urlprefix{URL }\fi
\providecommand{\bibinfo}[2]{#2}
\providecommand{\eprint}[2][]{\url{#2}}

\bibitem[{\citenamefont{Androsov et~al.}(2010)}]{telegin10}
\bibinfo{author}{\bibfnamefont{V.~P.} \bibnamefont{Androsov}}
  \bibnamefont{et~al.} (\bibinfo{year}{2010}), \bibinfo{note}{arXiv:1006.4846v1
  [physics.acc-ph]}.

\bibitem[{\citenamefont{Androsov et~al.}(2011)}]{telegin11}
\bibinfo{author}{\bibfnamefont{V.~P.} \bibnamefont{Androsov}}
  \bibnamefont{et~al.}, \bibinfo{journal}{PAST (Problems of Atomic Science and
  Technigue). Series: Nuclear Physics Investigations}
  \textbf{\bibinfo{volume}{55}}, \bibinfo{pages}{60} (\bibinfo{year}{2011}).

\bibitem[{\citenamefont{Androsov et~al.}(2005)}]{zelinsky05}
\bibinfo{author}{\bibfnamefont{V.~P.} \bibnamefont{Androsov}}
  \bibnamefont{et~al.}, \bibinfo{journal}{Nucl.Instrum. and Meth.A}
  \textbf{\bibinfo{volume}{543}}, \bibinfo{pages}{58} (\bibinfo{year}{2005}).

\bibitem[{CST()}]{CST}
\bibinfo{note}{CST Studio Suite$^{TM}$ 2010. http://www.cst.com}.

\bibitem[{\citenamefont{Cameron and Singh}(2009)}]{cameron09}
\bibinfo{author}{\bibfnamefont{P.}~\bibnamefont{Cameron}} \bibnamefont{and}
  \bibinfo{author}{\bibfnamefont{O.}~\bibnamefont{Singh}}, in
  \emph{\bibinfo{booktitle}{Proc. PAC-09 (Vancouver, Canada)}}
  (\bibinfo{year}{2009}), pp. \bibinfo{pages}{3392--3394}.

\bibitem[{\citenamefont{Cameron et~al.}(2009)\citenamefont{Cameron, Ferreira,
  and Krinsky}}]{cameron09a}
\bibinfo{author}{\bibfnamefont{P.}~\bibnamefont{Cameron}},
  \bibinfo{author}{\bibfnamefont{M.}~\bibnamefont{Ferreira}}, \bibnamefont{and}
  \bibinfo{author}{\bibfnamefont{S.}~\bibnamefont{Krinsky}}, in
  \emph{\bibinfo{booktitle}{Proc. PAC-09 (Vancouver, Canada)}}
  (\bibinfo{year}{2009}), pp. \bibinfo{pages}{4595--4597}.

\end{thebibliography}

\end{document}